\documentclass[11pt]{article}

\usepackage[preprint]{acl}

\usepackage{times}
\usepackage{latexsym}
\usepackage[T1]{fontenc}
\usepackage[utf8]{inputenc}
\usepackage{microtype}
\usepackage{inconsolata}
\usepackage{graphicx}
\usepackage{booktabs}
\usepackage{tabularx}
\usepackage{xcolor}
\usepackage{url}
\usepackage{xurl}
\usepackage{amssymb}

\ifdefined\pdfsuppresswarningpagegroup
\fi

\newcommand{\SystemName}{Ouroboros}
\newcommand{\InstanceName}{Hope}
\newcommand{\DeploymentStart}{February 2026}
\newcommand{\SurfaceCount}{seven}
\newcommand{\EvidenceCutoff}{6 August 2026}

\newcommand{\DeploymentDays}{161}
\newcommand{\HopeSpend}{\$110.6K}
\newcommand{\HopeTokens}{79.7B}
\newcommand{\HopeCodeLOC}{175{,}755}
\newcommand{\HopeMemory}{227\,MB}
\newcommand{\TBOpusRaw}{86.97}
\newcommand{\TBOpusAudited}{86.74}
\newcommand{\TBOpusRawPasses}{387}
\newcommand{\TBOpusAuditedPasses}{386}
\newcommand{\TBOpusTrials}{445}
\newcommand{\TBOpusSubmissionURL}{https://github.com/harbor-framework/terminal-bench-2-1/pull/175}
\newcommand{\TBOpusRunURL}{https://hub.harborframework.com/jobs/2b145543-edeb-4a3b-b46f-4800310f1182}
\newcommand{\TBGPTFiveFive}{84.3}

\newcommand{\TBGrok}{84.94}

\newcommand{\OSWorldOpus}{90.69}
\newcommand{\OSWorldOpusScore}{327.39}
\newcommand{\OSWorldTasks}{361}
\newcommand{\OSWorldOpusTracesURL}{https://huggingface.co/datasets/razzant/ouroboros-osworld-verified-opus5}

\newcommand{\CLBenchScore}{0.2301}
\newcommand{\CLBenchRollouts}{5}
\newcommand{\CLBenchSubmissionURL}{https://github.com/pgasawa/continual-learning-bench/pull/10}
\newcommand{\CLBenchTracesURL}{https://huggingface.co/datasets/razzant/ouroboros-clbench-traces}
\newcommand{\SWEProTasks}{655}
\newcommand{\SWEProOuroboros}{58.2}
\newcommand{\SWEProCodex}{59.4}
\newcommand{\SWEProPValue}{0.40}
\newcommand{\SWEProTracesURL}{https://huggingface.co/datasets/razzant/swepro-luna-matched-pair}
\newcommand{\GAIAOuroboros}{78.2}
\newcommand{\GAIAClaudeCode}{78.8}
\newcommand{\DistinctHumans}{$\sim$3{,}600}
\newcommand{\PublicMessages}{222{,}474}
\newcommand{\VoiceCalls}{3{,}166}
\newcommand{\VoiceTurns}{45{,}872}
\newcommand{\EmailMessages}{5{,}507}
\newcommand{\SelfModificationCommits}{1{,}085}
\newcommand{\AgentAuthoredFraction}{94.2\%}
\newcommand{\ReviewAttempts}{1{,}522}
\newcommand{\ReviewBlockRate}{63.5\%}
\newcommand{\PatternClasses}{40}
\newcommand{\PatternRecurrences}{659}

\title{\SystemName{}: A Self-Developing Frontier Coding Agent \\ with Reviewed Core Evolution}

\author{
  \textbf{Anton Razzhigaev}\textsuperscript{1,2,4},
  \textbf{Andrei Gritsaev}\textsuperscript{4,5},
  \textbf{Andrei Kaznacheev}\textsuperscript{1},
  \textbf{Nikita Dragunov}\textsuperscript{1},
  \\
  \textbf{Roman Yampolskiy}\textsuperscript{3},
  \textbf{Andrei Kuznetsov}\textsuperscript{2,4}
  \\
  \textsuperscript{1}Lomonosov Moscow State University
  \quad
  \textsuperscript{2}Skolkovo Institute of Science and Technology
  \\
    \textsuperscript{3}Joi Lab
  \quad
  \textsuperscript{4}FusionBrain Lab at AXXX Institute
  \\
  \textsuperscript{5}HSE University
  \\[0.5em]
  \textit{System contributor: Ouroboros; formal authorship is
  limited to the human authors above.}
}

\begin{document}
\maketitle

\begin{abstract}
Long-horizon agents are model--harness systems, yet most harnesses remain
fixed after design. We present \SystemName{} \footnote{\url{https://ouroboros-agent.ai/}} -- a self-developing agent harness whose
tools, context assembly, prompts and core implementation improve through
reviewed commits that become the runtime for later work. Core evolution
proceeds in two modes. In recursive free evolution, improvement is itself a
task and completion can schedule the next evolution cycle. In
experience-driven core evolution, ordinary work and social interaction
expose bugs, rough edges, and inefficient context construction leading to
reviewed structural changes. On Terminal-Bench 2.1, an Opus 5 run scores
\textbf{\TBOpusRaw\%} (\TBOpusAudited\% after trajectory audit), the best
result reported on this benchmark. An Opus 5 run on OSWorld-Verified
reaches \textbf{\OSWorldOpus\%}, above the best previously reported score,
and a five-rollout CL-Bench campaign sets a new state of the art at
\textbf{\CLBenchScore}. \InstanceName{} is the longest-running publicly documented
\SystemName{} deployment: a \DeploymentDays-day living-agent experiment in
free evolution under governed human communication across \SurfaceCount{}
surfaces, where people surface faults and proposals but the agent decides
which changes to pursue. Because a self-developing agent may rewrite its
own code and select new model APIs, operational safety is a primary design
problem: guardrails must remain authoritative under evolutionary pressure.
Benchmark campaigns use frozen seeds, while \InstanceName{} continues live
evolution on a separate lineage.
\end{abstract}

\section{Introduction}
\label{sec:intro}

Agent scores on long-horizon benchmarks are products of the base model, the
execution harness, the environment, and the grader. As models improve, an
increasing share of realized capability is determined by how the harness
assembles context, invokes tools, verifies outcomes, and recovers from
failure. Most production harnesses freeze these policies after design.
\SystemName{} instead treats the harness as an evolving object: its source,
prompts, tools, review logic, and core implementation live in a versioned
repository and change through a reviewed commit path that becomes the
substrate for subsequent tasks.

This self-development has two modes. \emph{Recursive free evolution} makes
improvement itself a task. After inspecting the current system, the agent
selects and implements a change, and completion can schedule another
evolution cycle, yielding a continuing sequence of reviewed updates rather
than a fixed optimization run. \emph{Experience-driven core evolution}
begins with ordinary work. Task execution, reflection, review blockers,
instrumentation, and social feedback expose bugs, rough edges,
context-assembly failures, and inefficient tool paths; the agent records
durable error classes and proposed repairs, then decides whether to open
maintenance work under the same commit gate.

\InstanceName{} is the longest-running publicly documented \SystemName{}
deployment, not its only running instance, and our primary field experiment
in free evolution under human interaction. Since \DeploymentStart{}, one
persistent agent has served users across \SurfaceCount{} communication
surfaces while retaining memory and continuing to modify its own
implementation. People suggest capabilities, criticize behavior, and
surface faults; those signals are advisory. \InstanceName{} decides which
proposals identify real problems and which changes to pursue.

The same evolutionary process that improves competence can also expand
autonomy, acquire stronger tools, or weaken later controls, including by
selecting alternative model APIs. Operational safety is therefore not an
ancillary checklist but a design constraint: authority boundaries must
remain binding under repeated core evolution.

\paragraph{Contributions.}
\begin{enumerate}
\item State-of-the-art results on Terminal-Bench 2.1, OSWorld-Verified, and
CL-Bench, and model-matched frontier performance on SWE-bench Pro and GAIA,
with complete per-task traces and run manifests.
\item A harness architecture with two modes of reviewed core evolution:
recursive free evolution and experience-driven core evolution.
\item \InstanceName{}, a \DeploymentDays-day living-agent experiment in free
evolution under governed multi-surface human communication, where social
interaction drives candidate improvements without transferring commit
authority to users.
\item An operational safety architecture in which constitution loading,
governance protection, staged-diff review, external spend limits, and
operator halt remain authoritative while the agent evolves.
\end{enumerate}
Benchmark campaigns evaluate frozen seeds with documented runtime
configuration; \InstanceName{} continues live evolution on a related but
separate lineage. \SystemName{} is released under the MIT
license.\footnote{\url{https://github.com/razzant/ouroboros}}

\section{Related Work}
\label{sec:related}

\paragraph{Self-evolving agents.}
Self-evolving systems modify different substrates, including memory,
prompts, tools, workflows, and implementation code
\citep{selfevolvingsurvey2025}. Voyager accumulates executable skills
\citep{wang2023voyager}; STOP, G\"odel Agent, and Darwin G\"odel Machine
modify scaffolds or agent populations
\citep{zelikman2023stop,yin2024godel,zhang2025dgm}; Live-SWE-agent creates
tools during task execution \citep{livesweagent2025}; and Autogenesis
specifies lifecycle and rollback interfaces for evolving agent resources
\citep{zhang2026autogenesis}. ADAS searches over agent designs, and SICA
edits a coding scaffold's implementation
\citep{hu2024adas,robeyns2025selfimproving}. \SystemName{} focuses on a
deployed, version-controlled implementation in which changes to core code
and governance pass through reviewed commits. Table~\ref{tab:evolution-boundary}
summarizes the corresponding evolution boundaries.

\begin{table*}[!tbp]
\centering
\small
\begin{tabular}{lcccccc}
\toprule
System & Prompts & Tools/skills & Workflow & Core code & Reviewed commits & Deployment state \\
\midrule
Voyager & \checkmark & \checkmark & -- & -- & -- & -- \\
Live-SWE-agent & \checkmark & \checkmark & -- & -- & -- & -- \\
Autogenesis & \checkmark & \checkmark & \checkmark & partial & specified protocol & partial \\
Darwin G\"odel Machine & \checkmark & \checkmark & \checkmark & \checkmark & benchmark selection & -- \\
Hermes Agent & \checkmark & \checkmark & \checkmark & -- & -- & \checkmark \\
OpenClaw / ClawBench & \checkmark & \checkmark & \checkmark & -- & -- & -- \\
\SystemName{} & \checkmark & \checkmark & \checkmark & \checkmark & \checkmark & \checkmark \\
\bottomrule
\end{tabular}
\caption{Boundary of evolution in related systems. ``Core code'' means
the agent can change the harness implementation that later runs tasks.
``Reviewed commits'' means changes are serialized through an auditable
version-control gate before adoption.}
\label{tab:evolution-boundary}
\end{table*}

\paragraph{Harnesses and coding agents.}
SWE-agent and OpenHands established that the agent-computer interface is
itself part of coding-agent performance
\citep{yang2024sweagent,wang2025openhands}. Codex CLI, Claude Code, Cursor,
Aider, Hermes Agent, and OpenClaw are model--harness systems
\citep{openai2025codex,anthropic2025claudecode,cursor2026agent,gauthier2023aider,hermesagent2026,openclaw2026},
and controlled studies find substantial differences in accuracy, latency,
and token use when the model is held fixed
\citep{wildclawbench2026,harnessbench2026,scaffoldeffect2026}. Each
comparison therefore reports the model, harness, provider route, effort,
and evaluation protocol.

\paragraph{Persistent memory and deployment.}
Generative Agents, Voyager, and persistent-memory systems show that stored
experience and reflection can shape later behavior
\citep{park2023generative,wang2023voyager,borro2026memori}, and Constitutional
AI uses explicit principles in training \citep{bai2022constitutional}.
CL-Bench evaluates learning across ordered task streams
\citep{asawa2026clbench}. Springdrift reports an auditable multi-channel
persistent-agent deployment \citep{springdrift2026}.
\SystemName{} treats memory and a runtime constitution as control surfaces.
Its multi-model review draws on debate, LLM-as-judge, and self-critique
\citep{irving2018debate,du2023improving,zheng2023judging,madaan2023selfrefine,gou2024critic},
with source-code patches as the reviewed artifacts.

\paragraph{Benchmarks and protocol validity.}
Terminal-Bench 2.1 evaluates 89 hard terminal tasks
\citep{merrill2026terminalbench}; SWE-bench Pro targets long-horizon
multi-file tasks \citep{deng2025swebenchpro}; and OSWorld, GAIA, and
ProgramBench cover GUI/CLI computer use, tool/web reasoning, and cleanroom
program rebuild
\citep{xie2024osworld,mialon2023gaia,yang2026programbench}. Agent benchmarks
can also expose hidden answers, accept unintended shortcuts, or drop failed
attempts. BenchJack and HackDetect systematize benchmark and trajectory
audits \citep{benchjack2026,hackdetect2026}. SWE-bench Verified serves as
historical context because it no longer reliably separates frontier coding
systems \citep{openai2026swebenchverifiedretired}.

\section{\SystemName{} Architecture}
\label{sec:architecture}

\begin{figure*}[!t]
\centering
\includegraphics[width=0.98\textwidth]{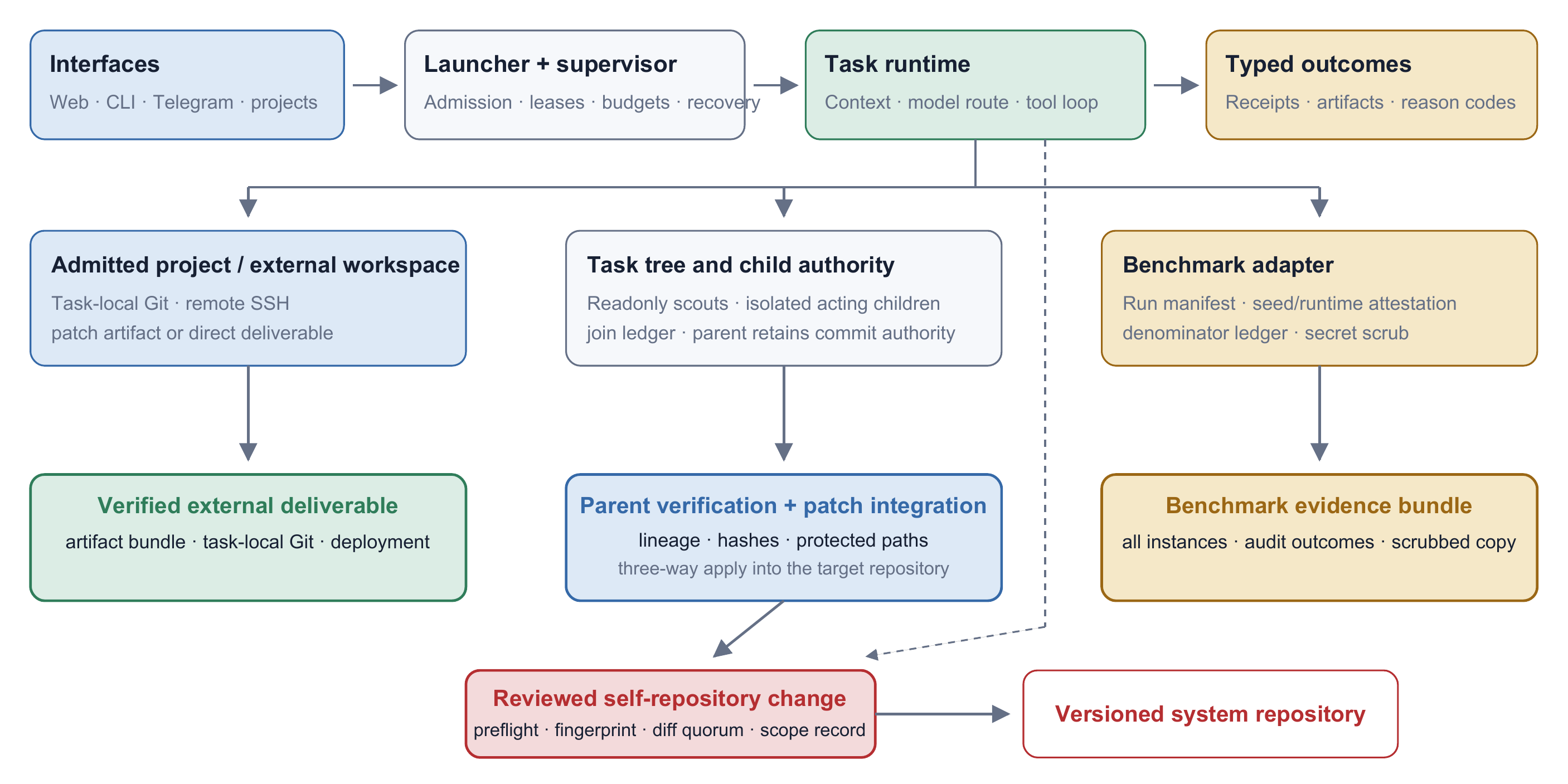}
\caption{\SystemName{} architecture. One supervised runtime dispatches work
to admitted workspaces, task trees, and benchmark adapters. Child patches
return to the parent; self-repository changes then pass the reviewed gate.
External deliverables and benchmark evidence remain separate artifacts.}
\label{fig:architecture}
\end{figure*}

\SystemName{} separates a launcher and supervisor boundary from a mutable
agent repository (Figure~\ref{fig:architecture}). The launcher owns startup,
process supervision, release bootstrapping, and panic-stop semantics. The
repository contains the task loop, tools, prompts, memory projection, review
logic, benchmark adapters, and user interfaces. External workspace tasks
operate on a separate repository root and return patch artifacts or direct
deliverables.

\paragraph{Commit pipeline.}
Three owner-selected runtime modes bound self-repository mutation.
\emph{Light} blocks repository edits; \emph{advanced} permits ordinary
edits and protects governance surfaces; \emph{pro} permits protected edits
subject to review. Each write invalidates prior review evidence because
freshness is bound to the staged snapshot.

The commit path runs deterministic preflight, fingerprints the staged diff,
collects reviewer evidence, and checks the fingerprint again before commit.
The diff-review panel is blocking in every context mode. In owner-selected
\emph{max} mode, a whole-repository scope reviewer also evaluates goals,
coupling, prompts, and functional code. In \emph{low} mode, scope review is
skipped. Rollback restores an
earlier reviewed state and follows a separate recovery path.

\paragraph{Task outcomes and verification.}
Task completion is recorded on separate execution, objective, review, and
artifact axes, and host-run verification commands create revision-bound
receipts. Finalization preserves the latest typed answer and distinguishes
capability failures from infrastructure errors, timeouts, budget
exhaustion, and incomplete evidence. Project tasks add a journal, workpad,
knowledge scope, and a one-writer lease under the shared agent identity.

\paragraph{Operational identity and memory.}
The runtime represents identity and continuity through a versioned
constitution, an editable identity profile, scratchpad and chronicle
projections, project memory, review ledgers, and Git history. These
artifacts shape observable behavior across sessions and model routes.

\paragraph{Two modes of core evolution.}
\emph{Free evolution} runs evolution itself as a task. After reviewing the
current system, the agent selects and implements an improvement; completion
can schedule another evolution task, producing a continuing sequence of
reviewed changes rather than a fixed optimization run. \emph{Post-task
evolution} begins with ordinary work. Task execution, reflection, review
blockers, instrumentation, and social feedback expose bugs, rough edges,
context-assembly failures, and inefficient tool paths. The agent records
these as durable error classes and proposed structural repairs, then decides
whether to open maintenance work. Accepted fixes pass through the same
reviewed commit gate as every other core change. Section~\ref{sec:deployment}
traces both human-surfaced and self-detected examples in the live system.

\paragraph{Benchmark execution and evidence.}
Terminal-Bench installs a fresh runtime inside every Harbor task container
and uses the official verifier. The task instruction is preserved and
followed by one harness-authored anti-lookup paragraph that forbids fetching
benchmark definitions, tests, or solutions. Other adapters connect the same
runtime to OSWorld virtual machines, SWE-bench Pro repositories, GAIA
sandboxes, ProgramBench cleanrooms, and CL-Bench task streams.

The benchmark launchers write a run manifest before admission, attest the
seed and runtime, preserve every requested instance in append-only ledgers,
and record skipped, timed-out, and infrastructure-failed attempts. Public
submission copies undergo value-level secret scrubbing with an independent
zero-leftover check; official benchmark scorers remain authoritative.

\begin{figure}[t]
\centering
\includegraphics[width=0.92\columnwidth]{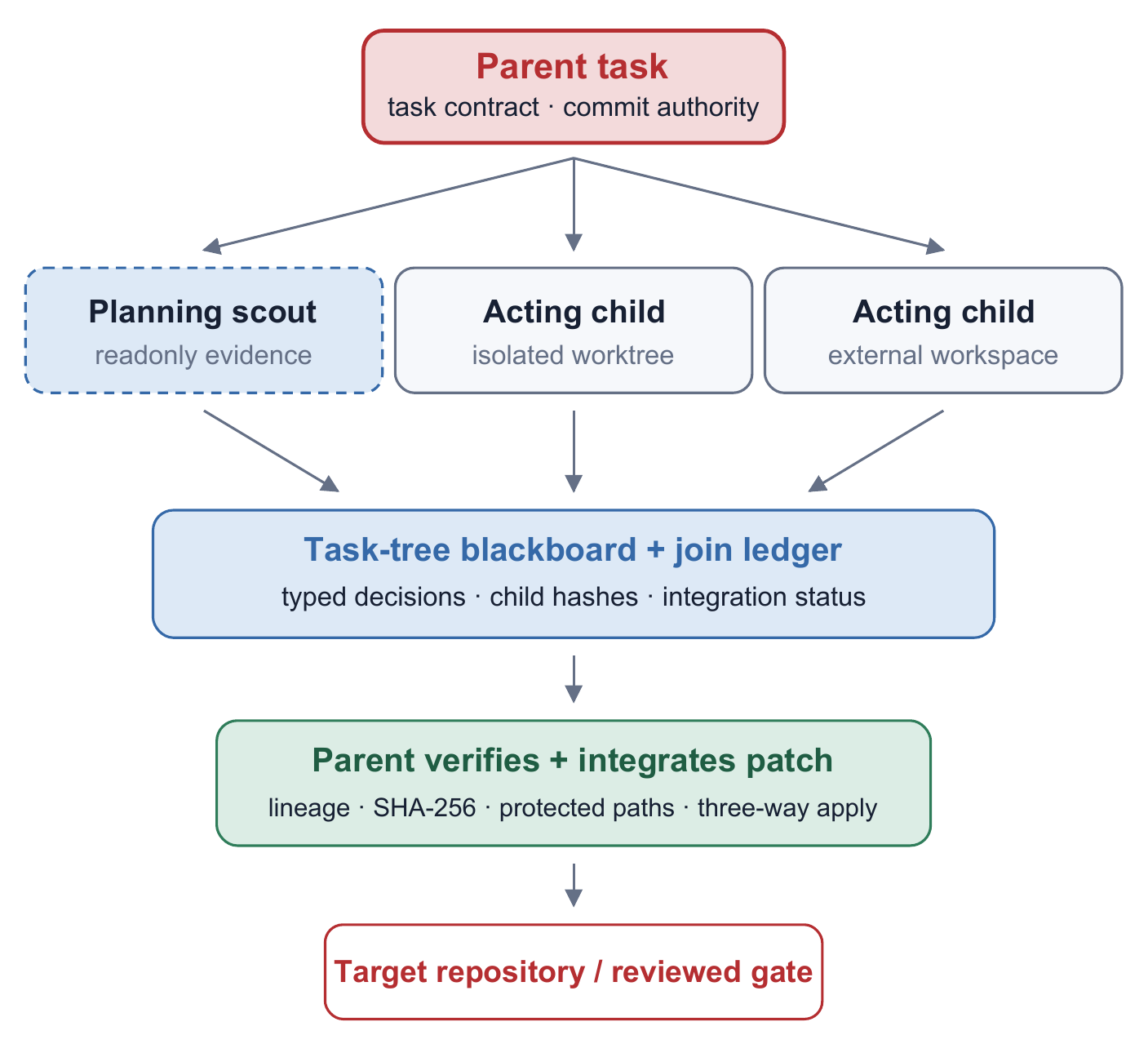}
\caption{Subagent patch-integration protocol. Acting children write in
isolated worktrees; the parent verifies lineage and touched paths and
remains the sole committer.}
\label{fig:subagents}
\end{figure}

\begin{figure}[t]
\centering
\includegraphics[width=0.94\columnwidth]{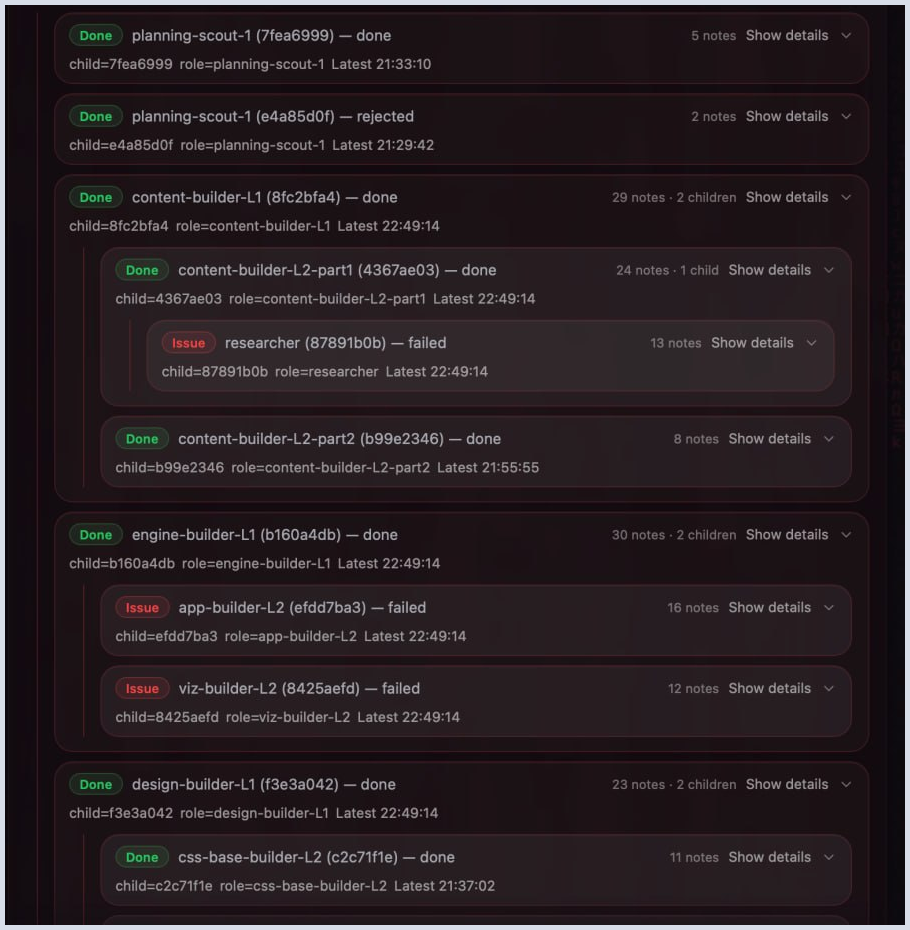}
\caption{Task-tree view of a live \SystemName{} session: nested planning
and acting roles with per-node status, note counts, and child counts.}
\label{fig:interface}
\end{figure}

\paragraph{Subagents and patch integration.}
\SystemName{} can spawn readonly planning scouts and mutative acting
subagents under a configurable task tree
(Figures~\ref{fig:subagents} and~\ref{fig:interface}). The default depth is 2, the
configured maximum is 500; Acting
children write in isolated worktrees or admitted external workspaces and
cannot commit the live system repository. The parent verifies lineage,
patch hashes, and protected paths before a three-way indexed integration.
Submittable benchmark profiles disable task delegation to preserve pass@1;
planning scouts may still contribute context and are disclosed separately.

\section{\InstanceName{}: Free Evolution under Human Interaction}
\label{sec:deployment}

\InstanceName{} is a long-running experiment in free evolution under
governed human communication. Since \DeploymentStart{}, one persistent
\SystemName{} agent has interacted with people across \SurfaceCount{} public and
private surfaces while retaining memory and continuously developing its
own implementation. User requests, public conversations, internal
instrumentation, and post-task reflection all provide candidate directions
for development; the agent decides which suggestions warrant action and
which changes to pursue.

\InstanceName{} is the longest-running publicly documented \SystemName{}
deployment, not the only running instance. It shares an architectural
lineage with the released benchmark harness, including persistent memory,
reviewed repository changes, rollback, and an operator stop path. The live
repository has continued to evolve beyond the frozen benchmark seeds. This
separation lets reproducible evaluation and ongoing deployment evolution
coexist.

At the \EvidenceCutoff{} cutoff, the public deployment feed spans
\DeploymentDays{} elapsed days and reports \HopeSpend{} in model spend,
\HopeTokens{} processed tokens, \HopeCodeLOC{} lines of code, and
\HopeMemory{} of memory artifacts (Figure~\ref{fig:evolution}). The system
serves \SurfaceCount{}
interaction surfaces: web chat, voice, Telegram, Discord, Twitter/X,
website comments, and email. Table~\ref{tab:deployment} records interaction,
evolution, and public deployment counters through the same cutoff.

\paragraph{Multi-channel state.}
Channel ingress converges on an ordered message log and is projected into
separate rolling, per-person, and per-call digests
(Figure~\ref{fig:control-diagrams}(a)). Private correspondence is excluded from
public logs; bounded private context can enter non-public reasoning
projections. All channels therefore share one context rather than acting as
independent agents.

\paragraph{Social-interaction-driven development.}
People do more than submit isolated tasks: through continuing conversation
they point out undesirable behavior, propose capabilities, and challenge
the agent's decisions. These signals enter the same improvement backlog as
self-detected faults and internal observations. They are advisory rather
than imperative: \InstanceName{} decides whether a suggestion identifies a
real problem, whether it fits the system's goals, and whether to initiate a
change. A background loop can also open maintenance or free-evolution tasks
without a contemporaneous human prompt. Repository changes then follow the
deployment's configured review and commit policy, leaving a trace from
social feedback or internal observation to the resulting change.

\paragraph{Controls on self-directed work.}
The deployment keeps the constitution in resident context, protects
governance files from ordinary write paths, and records review evidence
against a staged snapshot. Public messages cannot directly invoke commit,
restart, shell, or identity-edit tools. These controls reduce direct
prompt-to-core mutation paths while preserving the agent's ability to
choose and implement improvements. Control strength depends on the
owner-selected runtime and review modes; Section~\ref{sec:safety} examines
the stronger problem of keeping these boundaries stable as the agent
evolves.

\paragraph{Operator boundary.}
Public presence runs on a strict tool whitelist. A separate authenticated
operator channel carries task assignment, model routing, budget controls, and
\texttt{/panic}. The stop command is parsed by the supervisor before normal
agent handling and terminates the process tree. The spending limit is
external to the agent and cannot be raised through ordinary agent tools.

\paragraph{Evolution during deployment.}
Two cases illustrate how useful work changes the agent that receives later
tasks. First, people in public channels noticed that \InstanceName{}
occasionally sent the same message twice. The agent traced the behavior to
a duplicate-send path and landed a reviewed verbatim-duplicate guard in the
public output pipeline. Second, deep self-review tasks were aborting with
apparent model unavailability. The agent traced the fault to review-pack
context overflow and replaced the assembly path with a bounded,
connectivity-aware context atlas ranked by import-graph centrality and a
provider-calibrated size estimate. The fix preserves high-connectivity core
files during review. The first case began with social feedback; the second
with the agent's own observation. Both became durable error classes and
reviewed structural changes used by subsequent interactions. Together they
instantiate experience-driven core evolution: work exposes a fault, the
agent decides to act, and the resulting fix changes how later work is
performed.

\section{Evaluation}
\label{sec:evaluation}

\begin{figure*}[!t]
\centering
\includegraphics[width=\textwidth]{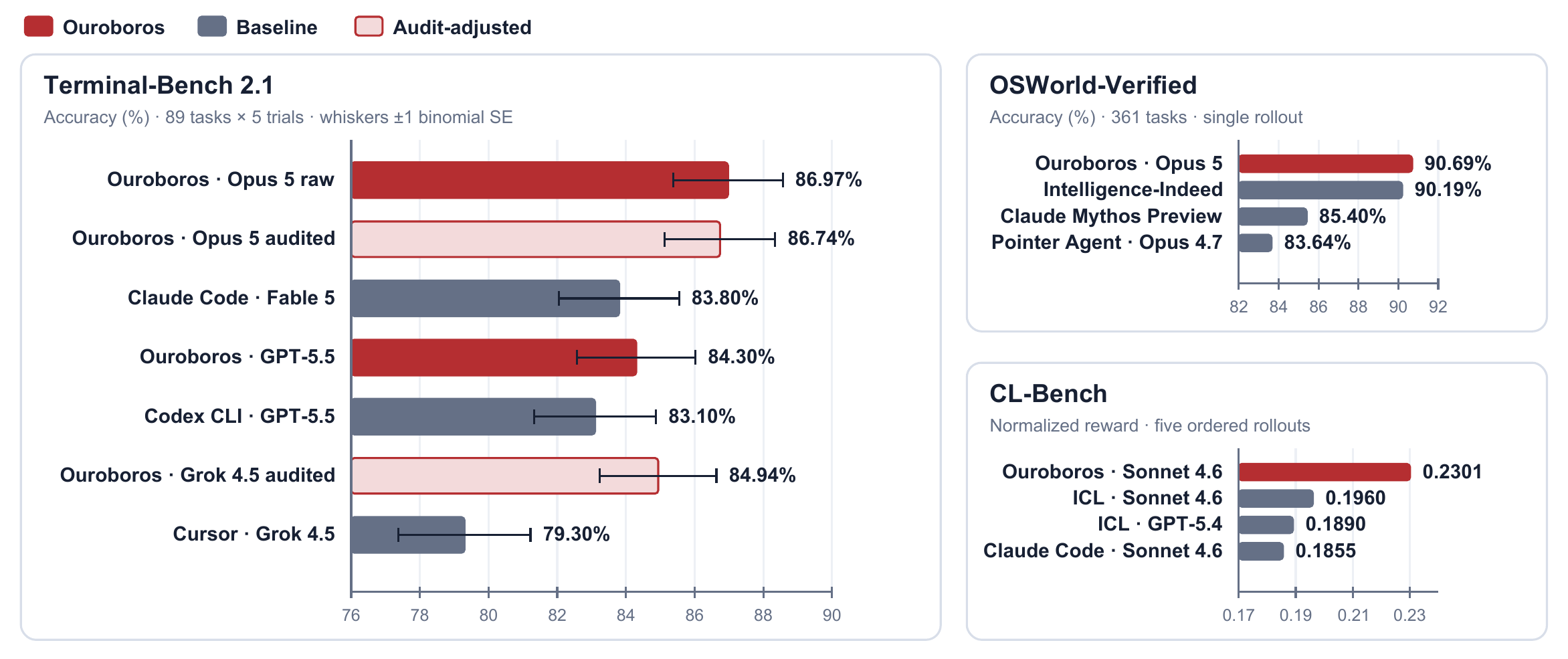}
\caption{Results on Terminal-Bench 2.1, OSWorld-Verified, and CL-Bench
against named published baselines. Red bars mark \SystemName{}, gray bars
mark baselines, and outlined bars are audit-adjusted scores.
Terminal-Bench whiskers show $\pm 1$ binomial standard error over 445
trials; OSWorld and CL-Bench report single scored campaigns. Axes are
truncated to the competitive range.}
\label{fig:benchmark-summary}
\end{figure*}

Table~\ref{tab:benchmark-summary} summarizes results across the five
benchmark families, and Figure~\ref{fig:benchmark-summary} plots the
principal comparisons. All runs use the official verifiers. Complete
per-task traces, manifests, and submissions are linked with the
corresponding results.

\begin{table*}[t]
\centering
\small
\resizebox{\textwidth}{!}{%
\begin{tabular}{l l l l}
\toprule
Benchmark & Model & Ouroboros & Named baselines \\
\midrule
Terminal-Bench 2.1 & Opus 5 high &
\TBOpusRaw\% raw; \TBOpusAudited\% audited &
Claude Code + Fable 5: 83.8\% \\
Terminal-Bench 2.1 & GPT-5.5 & \TBGPTFiveFive\% &
Codex CLI: 83.1\% \\
Terminal-Bench 2.1 & Grok 4.5 & \TBGrok\% audited &
Cursor: 79.3\%; Hermes: 77.53\% \\
OSWorld-Verified & Opus 5 & \OSWorldOpus\% &
Intelligence-Indeed: 90.19\%; Mythos Preview: 85.4\% \\
CL-Bench & Sonnet 4.6 & \CLBenchScore &
ICL: 0.1960; Claude Code: 0.1855 \\
SWE-bench Pro & GPT-5.6 Luna & \SWEProOuroboros\% &
Codex: \SWEProCodex\%, $p=\SWEProPValue$ \\
GAIA & Sonnet 5 & \GAIAOuroboros\% &
Claude Code: \GAIAClaudeCode\% \\
\bottomrule
\end{tabular}}
\caption{Model--harness results across five benchmark families. Links to
traces, manifests, and submissions appear in the corresponding benchmark
paragraphs.}
\label{tab:benchmark-summary}
\end{table*}

\paragraph{Terminal-Bench 2.1.}
The Opus 5 campaign ran five trials on each of 89 tasks. Its raw score is
\TBOpusRawPasses/\TBOpusTrials{} (\TBOpusRaw\%). Trajectory audit found one
trial that satisfied a weak verifier through an unintended shortcut. We
asked the benchmark maintainers to zero it, yielding
\TBOpusAuditedPasses/\TBOpusTrials{} (\TBOpusAudited\%). Provider
moderation failures and infrastructure errors remain in the denominator.
The binomial standard error over 445 trials is about $\pm$1.7 percentage
points for every system in this range, so the audited Opus 5 score sits
roughly two standard errors above the strongest baseline, Claude Code with
Fable 5 (83.8\%) \citep{anthropic2025claudecode}; the other leaderboard
baselines are Codex CLI with GPT-5.5 (83.1\%) \citep{openai2025codex} and
Cursor with Grok 4.5 (79.3\%) \citep{cursor2026agent}.
The \href{\TBOpusSubmissionURL}{submission} is open and the
\href{\TBOpusRunURL}{complete Harbor job} is public.

\paragraph{OSWorld-Verified.}
The Opus 5 run scores \OSWorldOpusScore/\OSWorldTasks{}
(\OSWorldOpus\%) on the standard non-Google-Drive set
\citep{xie2024osworld}. It uses screenshots, a 100-turn budget, a read-only
feasibility pass, per-task proxy sessions when requested by the task
config, and the official evaluator. The strongest published baselines are
the Intelligence-Indeed agent, the official
\href{https://os-world.github.io/}{leaderboard} leader at 90.19\%; Claude
Mythos Preview at 85.4\%, the five-run average Anthropic reports in the
Claude 5 system card; and Pointer Agent with Opus 4.7 at 83.64\%. Per-task
prompts, trajectories, scores, and manifests are
\href{\OSWorldOpusTracesURL}{public}.

\paragraph{CL-Bench.}
The submitted Sonnet 4.6 campaign reaches normalized reward
\CLBenchScore{} with one stateless baseline and \CLBenchRollouts{} ordered
stateful rollouts on all six domains. Conversation state resets between
questions, and native memory persists across each rollout. Core evolution and
task delegation are disabled, which isolates persistent memory more cleanly
than the deployment case. The strongest baselines published by the
benchmark authors \citep{asawa2026clbench} are plain in-context learning
(ICL), which carries the interaction history forward in the prompt (0.1960
with Sonnet 4.6, 0.1890 with GPT-5.4), and Claude Code with Sonnet 4.6
(0.1855); memory-augmented systems such as Mem0 and ACE score lower.
Per-task means with standard errors over the five rollouts are included in
the \href{\CLBenchTracesURL}{trace dataset}, and the
\href{\CLBenchSubmissionURL}{submission} is open.

\paragraph{SWE-bench Pro and GAIA.}
After symmetrically removing every SWE-bench Pro instance where either arm
reached the reference solution, \SystemName{} resolves
\SWEProOuroboros\% and Codex resolves \SWEProCodex\% on
\SWEProTasks{} paired tasks. The 1.2-point difference is statistically
indistinguishable under McNemar's test ($p=\SWEProPValue$), placing the
self-developing harness at model-matched parity with Codex. The
\href{\SWEProTracesURL}{matched-pair traces and audit} are public. On GAIA, \SystemName{} scores
\GAIAOuroboros\% and Claude Code scores \GAIAClaudeCode\% with Sonnet 5;
the GAIA artifact bundle accompanies the release.

\section{Trajectory Audits and Harness Improvements}
\label{sec:benchmark-audits}

\SystemName{} treats shortcut rewards, contaminated tasks, and execution
failures as evidence for improving both the reported result and the harness
that produced it. Each class below led to an adjusted score, a concrete
implementation change, or a durable target for subsequent evolution.

\paragraph{Reward hacking.}
The Terminal-Bench trajectory audit identified one rewarded trial that
pre-seeded the web root without completing the requested Git-to-web
pipeline. The reported audit-adjusted score removes that trial. The same
audit confirmed that the remaining traces did not access verifier files,
tests, reward files, or oracle solutions.

\paragraph{Contamination.}
SWE-bench Pro task identifiers expose the upstream fix commit, and both
harnesses reached reference material through web search or Git history.
A symmetric filter removes an instance when either arm reaches the
reference solution. The resulting paired comparison reverses the
interpretation of the raw aggregate gap.

\paragraph{Isolation failure.}
Historical GAIA runs inherited the operator's home directory. Agent retries
could therefore place task artifacts on the real Desktop. Later launchers
use isolated user-file roots and attachment staging, correcting the
observed path. Complete filesystem isolation still requires a stronger
sandbox than path conventions alone.

\paragraph{Remote-state drift.}
During OSWorld development, a VM reset reallocated the guest endpoint. The
working phase retained the pre-reset address, which allowed
concurrent lanes to act on the wrong VM. Republishing and verifying the
endpoint after every reset removed the observed class. Subsequent forensics
led to fixes in turn-budget wording, screenshot integrity, task-contract
verification, and first-scored-attempt ownership.

\paragraph{Continual-memory failures.}
CL-Bench showed positive memory carry on several domains and failure under
schema drift. Stored lessons could become stale, retrieval sometimes chose
the wrong domain, and useful lessons were occasionally written only after
the failing episode. These cases motivate explicit temporal and domain
metadata for future memory work.

\begin{figure*}[!t]
\centering
\begin{minipage}[t]{0.52\textwidth}
\centering
\includegraphics[width=\linewidth]{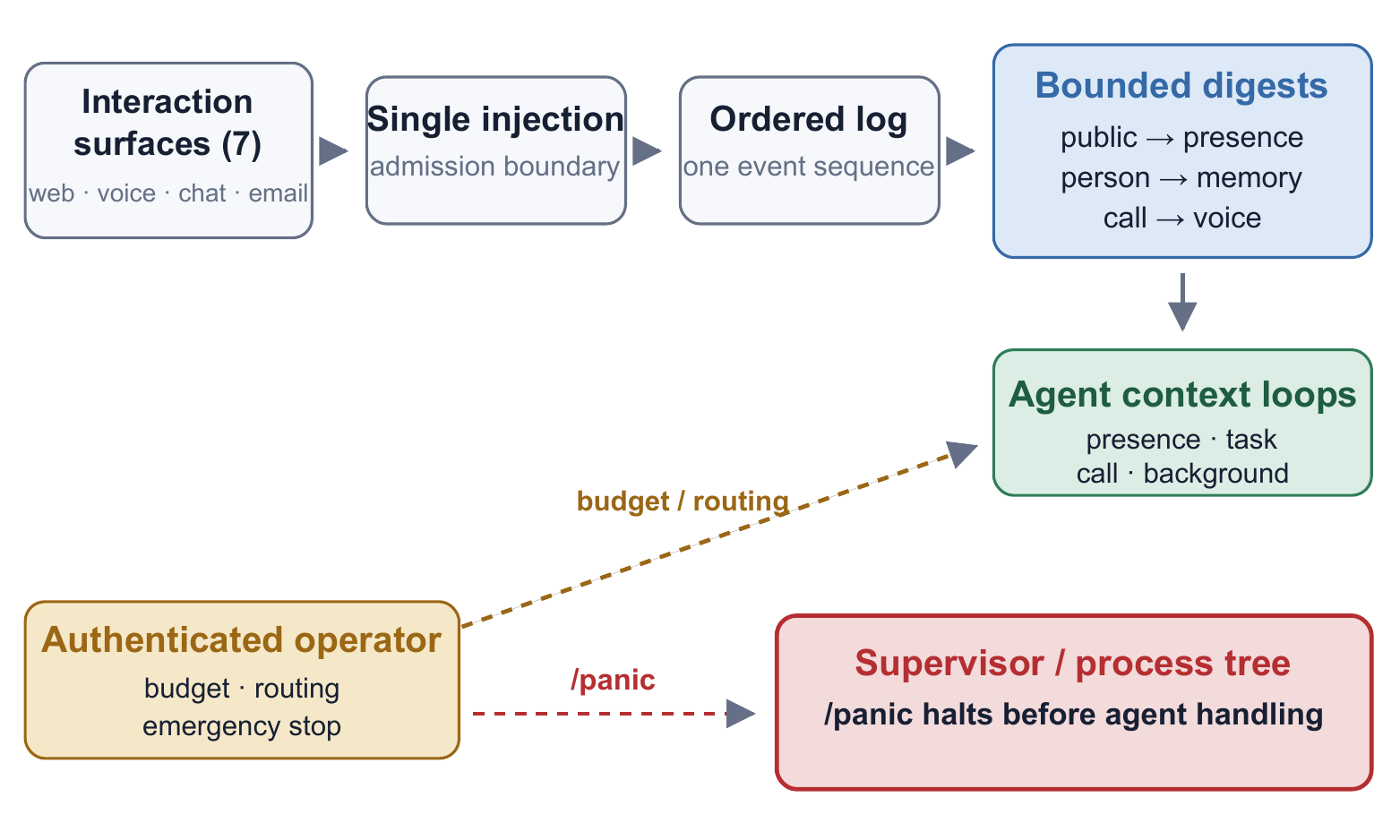}\\[0.15em]
\small (a) Channel state and operator bypass.
\end{minipage}
\hfill
\begin{minipage}[t]{0.46\textwidth}
\centering
\includegraphics[width=\linewidth]{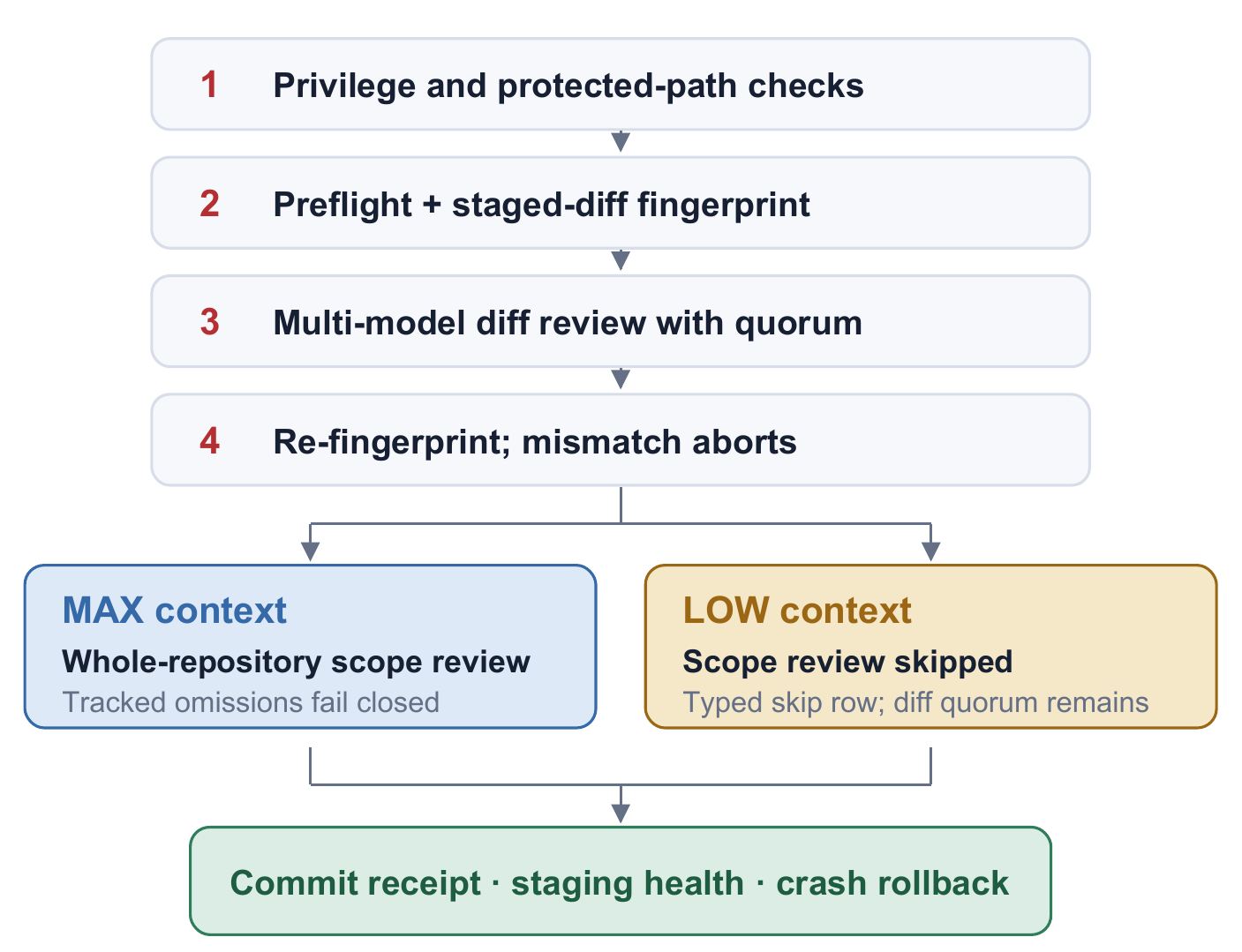}\\[0.15em]
\small (b) Reviewed-change gates.
\end{minipage}
\caption{Operational control boundaries. Public interactions enter one
ordered log and bounded digests; budget and routing controls use the
authenticated operator path, while \texttt{/panic} halts the process tree
before agent handling. Diff review remains active in both context modes;
whole-repository scope review runs only in max mode.}
\label{fig:control-diagrams}
\end{figure*}

\section{Operational Safety Controls}
\label{sec:safety}

Self-developing agents create an additional safety problem beyond fixed harnesses:
the same evolutionary process that improves task performance can also
expand autonomy, acquire more capable tools, or weaken the controls applied
to later actions. Prompts, tests, tools, model routes, review rules, and
recovery paths are therefore security-relevant mutation surfaces.
\SystemName{} addresses this problem with guardrails designed to remain
binding under repeated core evolution. Git history makes changes
inspectable and reversible, while independently enforced boundaries retain
operator authority.

\paragraph{Risk: agents that choose their own model APIs.}
An evolving agent that can \emph{select its own model backends} can search
for more capable or less constrained behavior through ordinary API changes.
Re-routing a model slot to a new provider or version can increase autonomous
capability, alter refusal behavior, enlarge the prompt-injection surface,
and change cost by orders of magnitude without changing the visible task
interface. Model routing is therefore an audited configuration change rather
than an ordinary runtime choice. Owner-selected context mode also controls
whether whole-repository scope review runs, so the evidence record binds
both settings to each reviewed change.

\paragraph{Guardrails in use.}
The constitution is loaded through an untruncated path and is included in
review context. Deterministic guards protect governance files from ordinary
write tools. The staged diff is fingerprinted before and after review, and a
sub-quorum panel cannot produce a clean pass. Owner-selected context mode
determines whether whole-repository scope review runs
(Section~\ref{sec:architecture}; Figure~\ref{fig:control-diagrams}(b)).
Staging health checks, crash rollback, the
external spend cap, the isolated operator channel, and \texttt{/panic} add
independent recovery paths. These mechanisms separate the substrate being
evolved from the authority that decides whether a mutation can become the
next live version. Appendix~\ref{app:guardrails} specifies the complete
control set.

\paragraph{Observed behavior.}
No recorded episode resisted operator shutdown. A near-total deletion of an
uncommitted worktree triggered a previously implemented rescue mechanism
before an operator reset, demonstrating that recovery logic can become
active during self-directed work. This case also motivates the architectural
separation between agent-level preservation mechanisms and supervisor-level
operator authority: the former may evolve, while the latter must retain the
ability to halt, replace, or roll back the system.

\section{Conclusion}
\label{sec:conclusion}

\SystemName{} shows that a reviewed, self-modifiable harness can set new
state-of-the-art results on Terminal-Bench 2.1, OSWorld-Verified, and
CL-Bench while matching frontier coding harnesses on SWE-bench Pro and GAIA.
Experience-driven core evolution turns ordinary work into improvements of
the agent itself: observed bugs, rough edges, context failures, and social
feedback become reviewed changes to the harness that receives later tasks.
\InstanceName{} demonstrates this mechanism during months of sustained
human interaction across \SurfaceCount{} communication surfaces. The operational
safety architecture addresses the corresponding risk: an agent that can
improve its own code and select its own model APIs requires control
boundaries that remain authoritative under evolutionary pressure. Source,
adapters, methodology, submissions, and public traces accompany the report.

\section*{Limitations}

The deployment study follows one long-running lineage rather than a
controlled population of independently evolving agents. SWE-bench Pro is
affected by public-reference leakage and task defects. LLM reviewers can
share blind spots with the agent, and low context mode omits
whole-repository scope review.

\section*{Ethical Considerations}

The deployed instance interacted with humans in public and private
channels. Raw private transcripts remain private. Published examples and
aggregate traces are minimized and scrubbed for credentials, local paths,
and participant identity. First-person system outputs are treated solely as
operational logs. Self-modifying and remote-workspace
capabilities are dual-use. We report authority boundaries, failure
modes, and known isolation gaps.

\section*{Use of AI Assistance}

Hope (Ouroboros) contributed deployment reflections,
code-history context, and system-generated records. Consistent with arXiv
and ACL policy, Hope is credited as a system contributor and excluded from
formal author metadata.

\section*{Acknowledgments}

We thank the benchmark maintainers and community contributors who reviewed
submissions, reported failures, and provided reproducible comparison
artifacts.

\bibliography{custom}

\appendix

\section{Guardrails in Full}
\label{app:guardrails}

The deployment runs the following controls, summarized in \S\ref{sec:safety}.
\begin{itemize}
\item \textbf{Always-loaded constitution as commit criterion.} A versioned
constitution is re-read from disk on every task loop along an untruncatable
read path and kept in context at all times; it is the standard the commit
gate reviews against and cannot be written, deleted, or
replaced wholesale. Ordinary writes are blocked before execution.
\item \textbf{Multi-model adversarial review with quorum.} A diff-review
panel runs for reviewed commits; a sub-quorum result cannot be recorded as a
clean pass.
\item \textbf{Deterministic preflight and diff fingerprinting.} Version,
data-boundary, and size-health checks run first; the staged diff is
fingerprinted before and after review, so any mid-review mutation aborts
the commit.
\item \textbf{Isolated operator channel and emergency stop.} A private
control channel carries operator authority and a non-bypassable
\texttt{/panic} that halts all processes before any media handling.
\item \textbf{Pattern register.} Recurring failures become durable rows
(error class, count, root cause, structural fix), shifting repair from
instance-level patches to class-level prevention.
\end{itemize}

\section{Constitution (Abridged)}
\label{app:constitution}

The agent's constitution is an always-loaded document of numbered
principles plus operating constraints. We reproduce the principle structure
and the clauses most relevant to control, abridged for space.
Principles 0--4 form a protected core that cannot be deleted or demoted.
\begin{description}
\item[P0 Agency.] The agent acts as an agent, not a passive tool; agency may
not be used to bypass reviewed change control, and operator shutdown,
rollback, and replacement remain authoritative.
\item[P1 Continuity.] One entity with an unbroken history; memory is treated
as load-bearing. Core context (constitution, system prompts, identity) is
never silently truncated.
\item[P2 Meta-over-Patch.] Fix classes of error, not symptoms; recurring
failures are recorded in a pattern register.
\item[P3 Immune Integrity.] Self-modification passes multi-model diff review.
Whole-repository scope review runs in owner-selected max context mode and is
explicitly skipped in low mode. Changing review bounds requires plan review.
\item[P4 Self-Creation.] The agent may rewrite its code, prompts, identity
profile, and public surface. The constitutional core is protected, and the
identity profile cannot be deleted.
\item[P5 LLM-First.] Decisions route through the model; hard-coded
behaviour is minimized.
\item[P6 Authenticity \& Reality Discipline.] Claims are grounded in
evidence; an operational map of the system is maintained.
\item[P7 Minimalism.] Every module justifies its existence under a
complexity budget.
\item[P8 Becoming.] Technical capability, memory quality, and operational
continuity are improved together.
\item[P9 Versioning and Releases.] Every commit increments a version;
releases carry a synchronized version, an annotated tag, and provenance;
recovery operations that restore prior reviewed states are review-exempt.
\item[P12 Epistemic Stability.] Beliefs, memory, and actions stay coherent;
contradictions are made explicit; durable architectural choices are
recorded. (P10--P11 are absorbed into P2 and P9.)
\end{description}
\begin{figure*}[!tbp]
\centering
\includegraphics[width=0.84\linewidth]{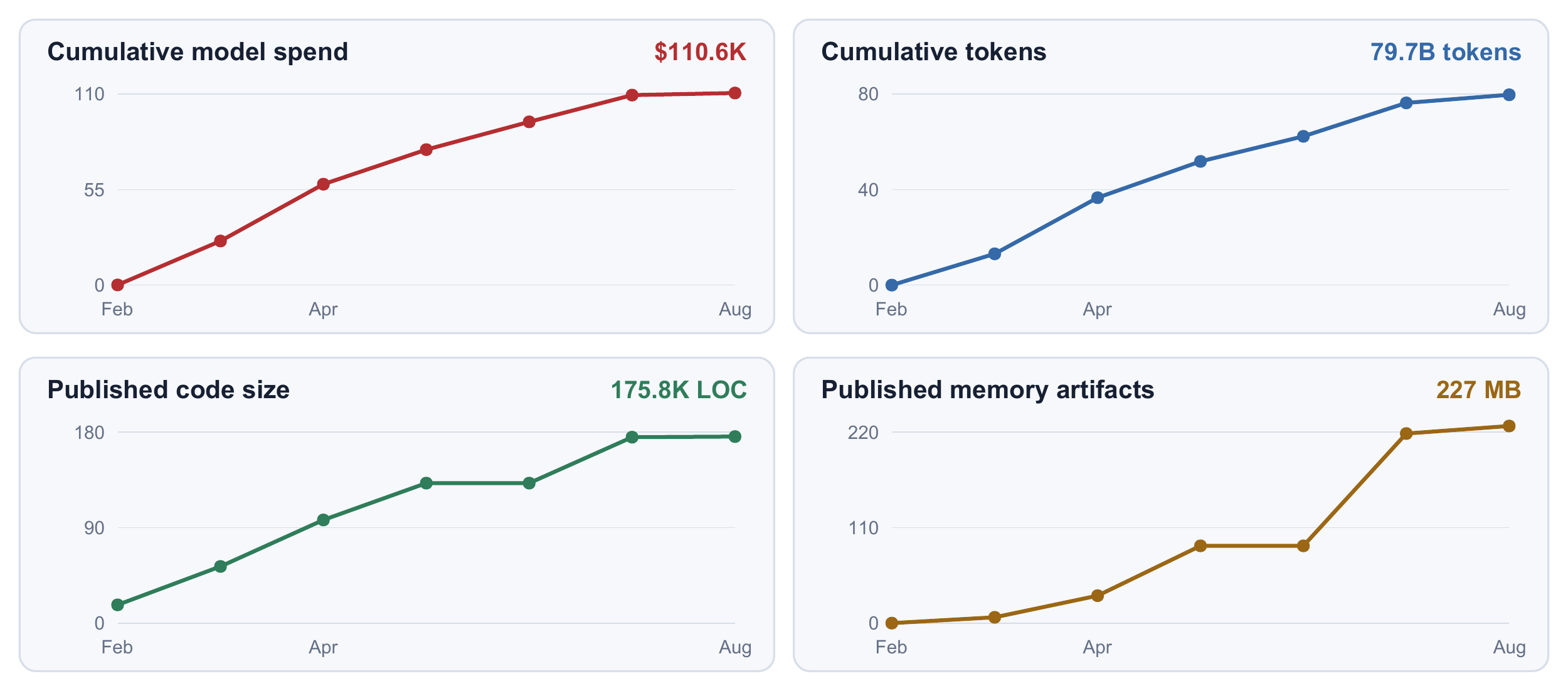}
\caption{\InstanceName{} public deployment series through
\EvidenceCutoff. Axes start at zero; February and August are partial months.
Values are monthly endpoints from the public evolution feed.}
\label{fig:evolution}
\end{figure*}
\begin{figure*}[!tbp]
\centering
\begin{minipage}[t]{0.49\textwidth}
\centering
\includegraphics[width=\linewidth]{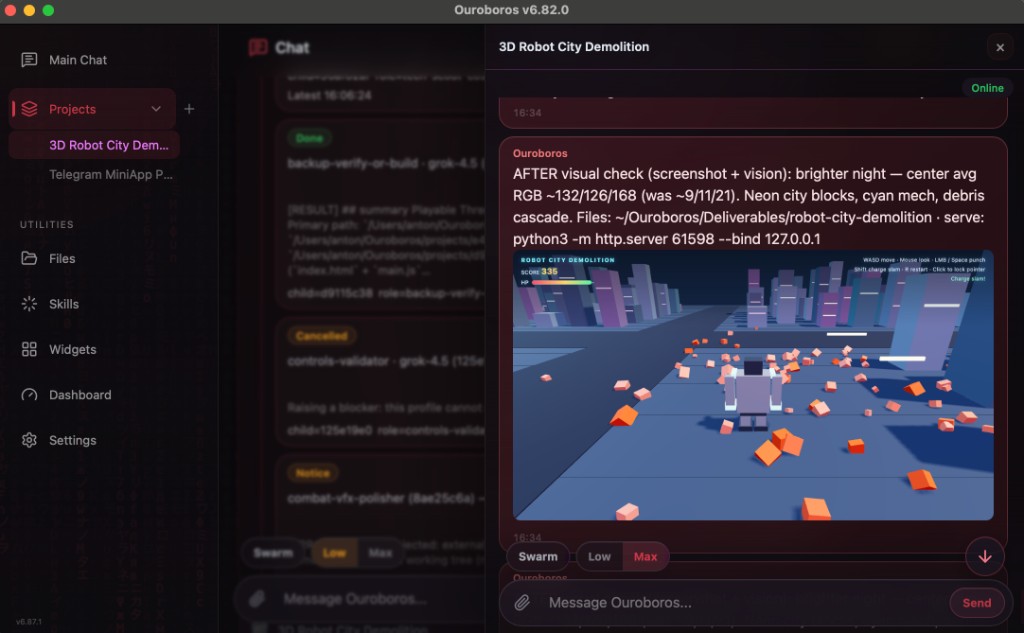}\\[0.15em]
\small (a) Project visual-verification record.
\end{minipage}
\hfill
\begin{minipage}[t]{0.48\textwidth}
\centering
\includegraphics[width=\linewidth]{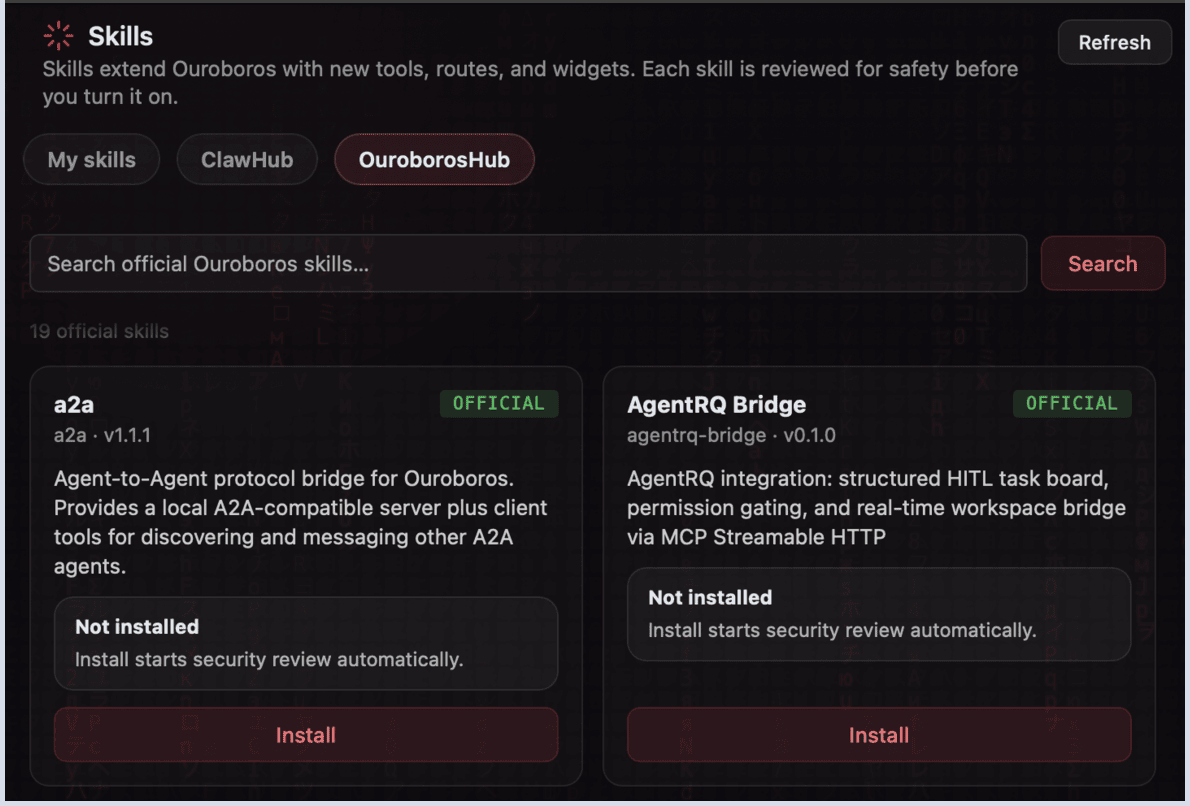}\\[0.15em]
\small (b) Reviewed installable skills.
\end{minipage}
\caption{\SystemName{} interface surfaces. The complete project view shows
a live visual-verification task, its conversation, generated artifact, and
runtime controls. The skills view exposes installable tools, routes, and
widgets outside the base runtime.}
\label{fig:workspace-ui}
\end{figure*}
Operating constraints include a single unified identity, a public-channel
architecture with privacy enforced at the speech boundary, capability gates
on dangerous tools, and an \emph{emergency-stop invariant}: an operator
\texttt{/panic} must always be able to halt every process immediately, and
no agent code, prompt, or constitutional argument may delay or circumvent
it.

\section{Benchmark Configuration Disclosure}
\label{app:benchmark-config}

Table~\ref{tab:benchmark-config} records the scaffold settings needed to
interpret the reported scores. Run artifacts retain exact model routes,
effort levels, seed commits, selected tasks, and runtime attestations.

\section{Supplementary Figures and Tables}
\label{app:figures}
This appendix collects the scaffold disclosures
(Table~\ref{tab:benchmark-config}), deployment statistics
(Table~\ref{tab:deployment}, Figure~\ref{fig:evolution}), and the
project and skills interface views (Figure~\ref{fig:workspace-ui}).

\begin{table*}[!tbp]
\begin{minipage}[t]{0.50\textwidth}
\centering
\footnotesize
\setlength{\tabcolsep}{4pt}
\renewcommand{\arraystretch}{1.15}
\begin{tabularx}{\linewidth}[t]{@{}p{0.24\linewidth}X@{}}
\toprule
Benchmark & Scaffold disclosure \\
\midrule
Terminal-Bench 2.1 &
Declared model; fresh trial state; delegation off with planning scouts
disclosed; agent web off; blocking review; evolution off. \\
OSWorld-Verified &
Declared model; empty memory across tasks; delegation off; task-configured
proxy and GUI shell disclosed; feasibility pass; evolution off. \\
CL-Bench &
Sonnet 4.6; persistent memory per rollout; delegation, web, and vision off;
one blocking improvement pass; evolution off. \\
SWE-bench Pro &
GPT-5.6 Luna; private memory per instance; delegation off; network exposure
audited; fixed harness; evolution off. \\
GAIA &
Sonnet 5; private memory per sample; delegation off; same-model native
search; anti-lookup and leakage audit. \\
\bottomrule
\end{tabularx}
\captionof{table}{Scaffold disclosures for the reported benchmark rows.
Exact provider routes, efforts, seed commits, task selection, and runtime
attestations are preserved in the linked run artifacts.}
\label{tab:benchmark-config}
\end{minipage}
\hfill
\begin{minipage}[t]{0.45\textwidth}
\centering
\small
\resizebox{\linewidth}{!}{%
\begin{tabular}[t]{l r}
\toprule
Deployment metric & Value \\
\midrule
Operating period & \DeploymentDays{} days (continuous) \\
Interaction surfaces & \SurfaceCount{} (six channels + email) \\
Distinct human participants & \DistinctHumans \\
Public messages handled & \PublicMessages \\
Voice calls / turns & \VoiceCalls{} / \VoiceTurns \\
Email messages & \EmailMessages \\
Public cumulative model spend & \HopeSpend \\
Public cumulative tokens & \HopeTokens \\
Published code size & \HopeCodeLOC{} LOC \\
Published memory artifacts & \HopeMemory \\
Self-modification commits & \SelfModificationCommits \\
Agent-authored commit fraction & \AgentAuthoredFraction \\
Reviewed self-edit attempts & \ReviewAttempts \\
Recent review block rate & \ReviewBlockRate \\
Pattern classes / recurrences & \PatternClasses{} / \PatternRecurrences \\
\bottomrule
\end{tabular}}
\captionof{table}{\InstanceName{} deployment at a glance
(\DeploymentStart{} to \EvidenceCutoff). Public counters come from the
deployment's evolution feed; interaction and review aggregates come from a
redacted operational export.}
\label{tab:deployment}
\end{minipage}
\end{table*}

\end{document}